\documentclass[preprint,showpacs,preprintnumbers,amsmath,amssymb]{revtex4}
\usepackage{amsmath}

\usepackage{graphicx}
\usepackage{pdfpages}
\usepackage{dcolumn}
\usepackage{bm}
\usepackage{amsmath} 
\usepackage{amsfonts}
\usepackage{amssymb}
\usepackage{url}
\usepackage{microtype}
\usepackage{graphicx}
\usepackage{appendix}%

\newcommand{\qed}{\nobreak \ifvmode \relax \else
      \ifdim\lastskip<1.5em \hskip-\lastskip
      \hskip1.5em plus0em minus0.5em \fi \nobreak
      \vrule height0.75em width0.5em depth0.25em\fi}

\begin{document}

\preprint{}
\title{Archipelagos of Total Bound and Free Entanglement}
\author{Paul B. Slater}
 \email{slater@kitp.ucsb.edu}
\affiliation{%
Kavli Institute for Theoretical Physics, University of California, Santa Barbara, CA 93106-4030\\
}
\date{\today}
            
\begin{abstract}
First, we considerably simplify an initially quite complicated formula--involving dilogarithms. It yields the total bound entanglement probability 
($\approx 0.0865542$) for a  qubit-ququart ($2 \times 4$) three-parameter model, recently analyzed for its separability properties by Li and Qiao. An ``archipelago" of disjoint bound-entangled regions appears in the space of parameters, somewhat similarly to those recently found in our preprint, ``Jagged Islands of Bound Entanglement and Witness-Parameterized Probabilities". There, two-qutrit and two-ququart  Hiesmayr-L{\"o}ffler ``magic simplices" and generalized Horodecki states had been examined. However, contrastingly, in the present study, the {\it entirety} of bound entanglement--given by the formula obtained--is clearly captured in the archipelago found. Further, we ``upgrade'' the qubit-ququart model to a two-ququart one, for which we again find a bound-entangled  archipelago, with its total probability simply  being now $\frac{1}{729} \left(473-512 \log \left(\frac{27}{16}\right) \left(1+\log
   \left(\frac{27}{16}\right)\right)\right) \approx 0.0890496$. Then, ``downgrading" the qubit-ququart model to  a two-qubit one, we  find an archipelago of total  {\it non-bound/free} entanglement probability $\frac{1}{2}$. 
\end{abstract}
 

\maketitle
The question of whether any particular quantum state is separable (``classically correlated" \cite{werner1989quantum}) or entangled (``EPR-correlated") is, in general, a highly challenging (NP-hard) one to address \cite{gharibian2008strong,akulin2015essentially,de2018symplectic}. Only in dimensions $2 \times 2$ and $2 \times 3$ is a simple
(positive-partial-transpose [PPT]) both necessary and sufficient condition at hand (cf. \cite{sperling2009necessary}). Higher-dimensional states that are not separable, but are nonetheless PPT are designated as ``bound-entangled" \cite{sanpera2001schmidt}, a phenomenon of widespread/foundational interest (e. g. \cite{tendick2019activation}). An interesting (geometric/topological/disjointed \cite{slater2019bound}) aspect of this will be presented here in multiple examples, based on those put forth by  Li and  Qiao in sec. 2.3.1 of their recent paper, ``Separable Decompositions of Bipartite Mixed States" \cite{li2018separable}. Within their framework, one can definitively conclude whether any specific state is separable or not.
(Somewhat contrastingly, Gabdulin and Manilara employ {\it numerical} 
methods--the best separable approximation--for such identification purposes in the case of two-qutrit states \cite{gabdulin2019investigating}.)

The first example of Li and Qiao we use is that of the $2 \times 4$ dimensional mixed (qubit-ququart) state,
\begin{equation} \label{rhoAB}
\rho_{AB}^{(1)}=\frac{1}{2 \cdot 4} \textbf{1} \otimes \textbf{1} +\frac{1}{4} (t_1 \sigma_1 \otimes \lambda_1+t_2 \sigma_2 \otimes \lambda_{13}+t_3 \sigma_3 \otimes \lambda_3),
\end{equation}
where $t_{\mu} \neq 0$, $t_{\mu} \in \mathbb{R}$, and $\sigma_i$ and $\lambda_{\nu}$ are SU(2) (Pauli matrix) and SU(4) generators, respectively (cf. \cite{singh2019experimental}).

\begin{figure}
    \centering
    \includegraphics{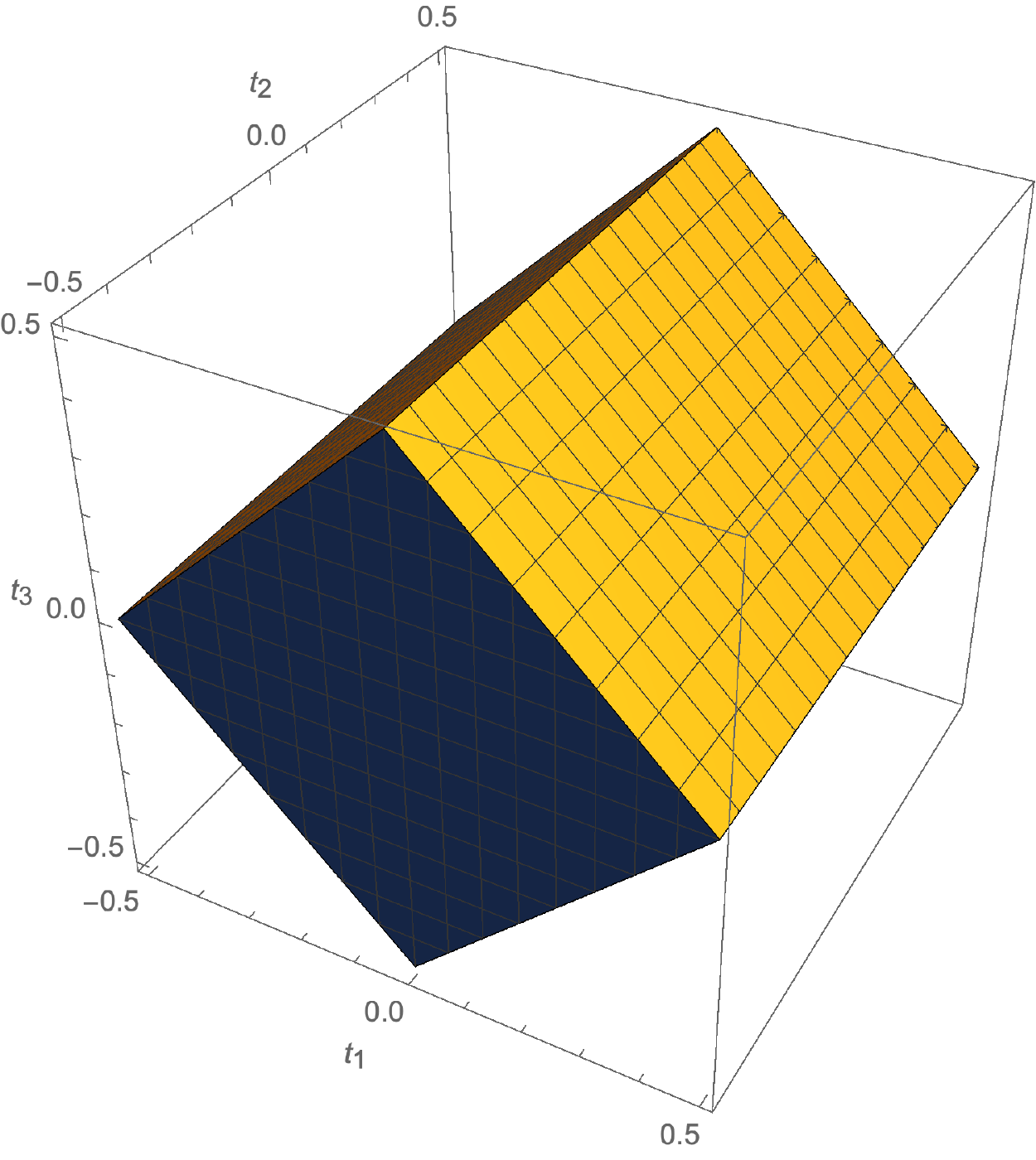}
    \caption{The convex set--in accordance with the constraints (\ref{eq2})--of possible physical states representable by $\rho_{AB}^{(1)}$, given by (\ref{rhoAB}).}
    \label{fig:LiQiaoPPT}
\end{figure}

They asserted that equation (\ref{rhoAB}) represents a physical state when the $8 \times 8$ density matrix $\rho_{AB}^{(1)}$ is positive semidefinite, that is when
\begin{equation} \label{eq2}
t_2 \leq \frac{1}{4}, \hspace{.1in} (|t_1|+|t_3|)^2 \leq 
\frac{1}{4}.
\end{equation}
They also found that  $\rho_{AB}^{(1)}$
 has positive (semidefinite) partial transposition, so the well-known PPT criterion cannot be used to help determine whether any specific state is entangled or separable.
Figure~\ref{fig:LiQiaoPPT} shows the convex set of possible physical states representable by $\rho_{AB}^{(1)}$.

After an extended analysis, using the interesting tools they develop, Li and Qiao reach the conclusion \cite[eq. (59)]{li2018separable} that $\rho_{AB}$ is entangled when 
\begin{equation} \label{eq3}
  (|t_1|+|t_2|+|t_3|)^2>1 \hspace{.1in}  \mbox{or} \hspace{.1in} (t_1 t_2 t_3)^2 > \frac{1}{27} \cdot \Big(\frac{2}{27}\Big)^2 =\frac{4}{27^3},
\end{equation}
where the quantity $\frac{1}{27}$ is associated with the qubit and the $\Big(\frac{2}{27}\Big)^2$ with the ququart. We will later again utilize two of these quantities/lower-bounds in related two-ququart and two-qubit analyses. Li and Qiao stated \cite[eqs. (57), (58)]{li2018separable} that the first bound ($\frac{1}{27}$) can be associated with the states
\begin{equation}
\rho^{(1)(A)} = \frac{1}{2} \textbf{1} + \frac{1}{2\sqrt{3}}( \pm \sigma_1 \pm \sigma_2 \pm \sigma_3),
\end{equation}
and the second $\Big(\frac{2}{27}\Big)^2$ with the states
\begin{equation}
\rho^{(1)(B)} = \frac{1}{4}\textbf{1} + \frac{1}{2}(\pm \frac{ \sqrt{2}}{3} \lambda_1 \pm \frac{\sqrt{2}}{3} \lambda_3 + \frac{1}{3}\lambda_{13} + \frac{1}{3\sqrt{3}}\lambda_{8} + \frac{1}{3\sqrt{6}}\lambda_{15} )  .  
\end{equation}
(It appears that the derivation of these states could be further clarified, and their counterparts and associated bounds explicitly given in the pair of two-qutrit models also studied by Li and Qiao.)

Let us now--to proceed in a probabilistic framework--standardize (dividing by one-half) the three-dimensional Euclidean volume of the possible physical states of $\rho_{AB}^{(1)}$ (Fig.~\ref{fig:LiQiaoPPT}) to equal 1. Then, if we impose the  entanglement constraint $ (t_1 t_2 t_3)^2 >\frac{4}{27^3}$ given above, we  obtain--taking into account the noted PPT property of $\rho_{AB}^{(1)}$--the (Hilbert-Schmidt) probability of {\it bound} entanglement. (The other constraint, $(|t_1|+|t_2|+|t_3|)^2>1$, in (\ref{eq3}), itself proves to be unrealizable/unenforceable, and not of any relevance to this entanglement calculation.)

This bound entanglement probability ($P$) for the Li-Qiao qubit-ququart model (\ref{rhoAB}) takes--at least upon first analysis/constrained-integration--the quite involved  form 
\begin{equation} \label{bound}
P=  \frac{9 \sqrt{243-64 \sqrt{3}}-4 \Big(16 \coth ^{-1}\left(\frac{9}{\sqrt{81-\frac{64}{\sqrt{3}}}}\right)+A+B+C\Big)}{81 \sqrt{3}}  \approx 0.08655423366978987,
\end{equation}
where
\begin{displaymath}
A=2 \log \left(\frac{1024}{243} \left(9+\sqrt{81-\frac{64}{\sqrt{3}}}\right)\right) \log
   \left(27-\sqrt{729-192 \sqrt{3}}\right)-3 \log (48) \log (108)
\end{displaymath}
and
\begin{displaymath}
B=2 \log ^2\left(27+\sqrt{729-192 \sqrt{3}}\right)+3 \log \left(\frac{2187}{256}\right)
   \log \left(27+\sqrt{729-192 \sqrt{3}}\right)
\end{displaymath}
and, with the polylogarithmic function being employed,
\begin{displaymath}
C=8 \text{Li}_2\left(\frac{1}{18} \left(9-\sqrt{81-\frac{64}{\sqrt{3}}}\right)\right)-8
   \text{Li}_2\left(\frac{1}{18} \left(9+\sqrt{81-\frac{64}{\sqrt{3}}}\right)\right).
\end{displaymath}
 We interestingly noted that only 2's and 3's occur in the prime decompositions of the several integers present in the formula, and also that 
 $\sqrt{729-192 \sqrt{3}}=3 \sqrt{81-\frac{64}{\sqrt{3}}}$.
Also observed (by  Mathematica) was that
\begin{equation}
\log \left(27+\sqrt{729-192 \sqrt{3}}\right)-\log \left(27-\sqrt{729-192 \sqrt{3}}\right)=2 \coth ^{-1}\left(\frac{9}{\sqrt{81-\frac{64}{\sqrt{3}}}}\right)
\end{equation}
and
\begin{equation}
\log \left(27+\sqrt{729-192 \sqrt{3}}\right)-\coth
   ^{-1}\left(\frac{9}{\sqrt{81-\frac{64}{\sqrt{3}}}}\right)=3 \log (2)+\frac{3 \log (3)}{4}.
\end{equation}

Making use of these last two identities, we arrived at the considerably simpler  formula for the bound entanglement probability,
\begin{equation} \label{finalformula}
P=\frac{16 (-4-9 \log (3)+8 \log (2)) \coth ^{-1}\left(\frac{9}{\sqrt{81-\frac{64}{\sqrt{3}}}}\right)}{81 \sqrt{3}} +
\end{equation}
\begin{displaymath}
\frac{32 \left(\text{Li}_2\left(\frac{1}{18}
   \left(9+\sqrt{81-\frac{64}{\sqrt{3}}}\right)\right)-\text{Li}_2\left(\frac{1}{18}
   \left(9-\sqrt{81-\frac{64}{\sqrt{3}}}\right)\right)\right)+9 \sqrt{3} \sqrt{81-\frac{64}{\sqrt{3}}}}{81 \sqrt{3}}.
\end{displaymath}
The polylogarithms (dilogarithms) remain, however,  as in the original formula.

In such regards, we have further observed as part of a simplification analysis  \cite{Slater19Formula} --changing the subscript of Li from 2 to 1 (leading to the standard logarithmic framework)--that 
\begin{equation}
 \text{Li}_1\left(\frac{1}{18}
   \left(9+\sqrt{81-\frac{64}{\sqrt{3}}}\right)\right)-\text{Li}_1\left(\frac{1}{18}
   \left(9-\sqrt{81-\frac{64}{\sqrt{3}}}\right)\right) = 
\end{equation}
\begin{equation}
\log \left(\frac{1}{18} \left(9+\sqrt{81-\frac{64}{\sqrt{3}}}\right)\right)-\log
   \left(\frac{1}{18} \left(9-\sqrt{81-\frac{64}{\sqrt{3}}}\right)\right)=
\end{equation}
\begin{equation}
2 \coth ^{-1}\left(\frac{9}{\sqrt{81-\frac{64}{\sqrt{3}}}}\right),
\end{equation}

Carlo Beenakker, then, observed that 
\begin{equation}
 \text{Li}_2\left(\frac{1}{18}
   \left(9+\sqrt{81-\frac{64}{\sqrt{3}}}\right)\right)-\text{Li}_2\left(\frac{1}{18}
   \left(9-\sqrt{81-\frac{64}{\sqrt{3}}}\right)\right)= 
\end{equation}
\begin{displaymath}
\int_{\frac{1}{2}-\frac{1}{18}
   (9-\sqrt{81-\frac{64}{\sqrt{3}}}}^{\frac{1}{2}+\frac{1}{18}
   (9-\sqrt{81+\frac{64}{\sqrt{3}}}} \frac{\log{1-t}}{dt}.
\end{displaymath}
He wrote ``an answer in terms of elementary functions is unlikely; and an answer in terms of special functions is what you have" \cite{Slater19Formula2}, thus, indicating that no further simplification is achievable.

Let us note that the difference of two dilogarithmic functions is present--as well as inverse hyperbolic tangents--in an auxiliary formula of Lovas-Andai for the Hilbert-Schmidt separability probability ($\frac{29}{64}$) of the two-rebit states,
\begin{equation} \label{poly}
\tilde{\chi}_1 (\varepsilon)=\frac{2 \left(\varepsilon ^2 \left(4 \text{Li}_2(\varepsilon )-\text{Li}_2\left(\varepsilon
   ^2\right)\right)+\varepsilon ^4 \left(-\tanh ^{-1}(\varepsilon )\right)+\varepsilon ^3-\varepsilon
   +\tanh ^{-1}(\varepsilon )\right)}{\pi ^2 \varepsilon ^2},    
\end{equation}
\cite[eq. (2)]{slater2018master} \cite[eq. (9)]{lovas2017invariance}, where $\varepsilon$ is a ratio of singular values. Further, this is the specific case (random matrix Dyson index) $d=1$ of a ``master Lovas-Andai formula" \cite[eq. (70)]{slater2018master} 
\begin{equation}
\tilde{\chi}_d (\varepsilon)=\  \frac{\varepsilon^d \Gamma(d+1)^3 \, _3\tilde{F}_2\left(-\frac{d}{2},\frac{d}{2},d;\frac{d}{2}+1,\frac{3 d}{2}+1;u\right)}{\Gamma(\frac{d}{2}+1)^2}. 
\end{equation}
The regularized hypergeometric function is indicated, with the case $d=2$ corresponding to the standard two-qubit scenario, for which the evidence is strongly compelling that the associated Hilbert-Schmidt separability probability is $\frac{8}{33}$ \cite{slater2019numerical}.

\begin{figure}
    \centering
    \includegraphics{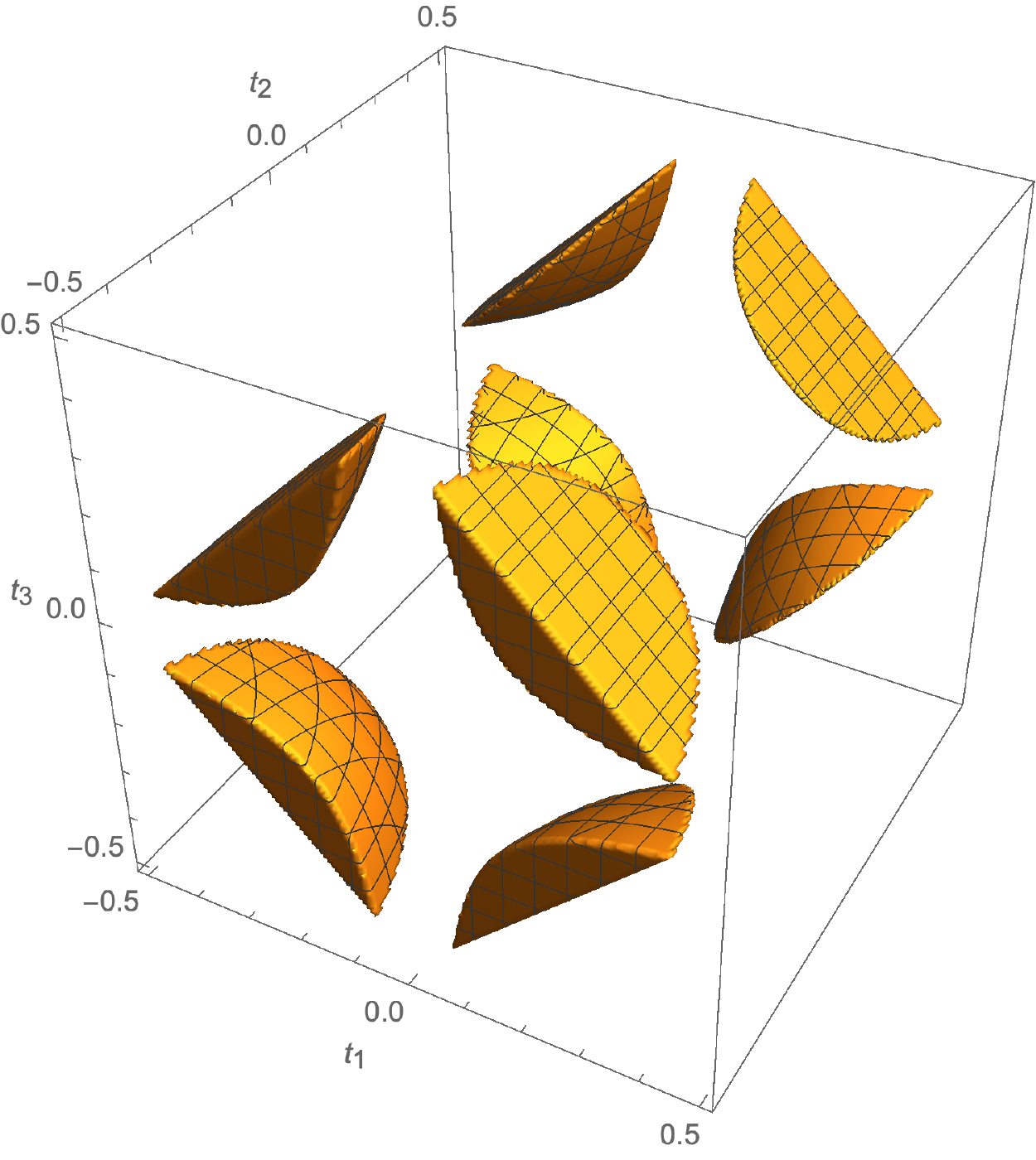}
    \caption{Bound-entangled archipelago for the Li-Qiao 
    $2 \times 4$ dimensional mixed (qubit-ququart) state $\rho_{AB}$  (\ref{rhoAB}). The total probability ($\approx 0.0865542$) captured by the islands is given by eq. (\ref{finalformula}).}
    \label{fig:LiQiao}
\end{figure}

In Fig.~\ref{fig:LiQiao}, we show the {\it archipelago}  determined by the joint enforcement of the physicality constraints (\ref{eq2}) and the multiplicative one of (\ref{eq3}), for which the bound entanglement probability assumes the indicated value ((\ref{bound}), (\ref{finalformula})).
(It would be an interesting exercise to formally enumerate/delimit the total number of such islands--eight appearing in the figure.)

Similar nonsmooth behavior of disjointed  regions of bound entanglement was reported in our recent preprint \cite{slater2019bound}, ``Jagged Islands of Bound Entanglement and Witness-Parameterized Probabilities" 
(cf. \cite{gabdulin2019investigating}). There, (two-qutrit and two-ququart) Hiesmayr-L{\"o}ffler ``magic simplices" \cite{hiesmayr2014mutually,baumgartner2007special} and generalized Horodecki states \cite{horodecki1998mixed} had been analyzed. (The several bound-entanglement probabilities obtained there--such as $-\frac{4}{9}+\frac{4 \pi }{27 \sqrt{3}}+\frac{\log (3)}{6} \approx 0.00736862$--were a magnitude smaller than the 0.08655423 reported here. Also, several highly challenging to obtain witness-parameterized families of bound-entangled probabilities were reported.) 

Here, we have been able--in the Li-Qiao framework--to 
account for the {\it entirety} of bound entanglement, while in the earlier indicated 
study \cite{slater2019bound}--making use  of entanglement witnesses, the computable cross-norm or realignment (CCNR) and SIC-POVM (symmetric informationally complete positive operator-valued measure) 
criteria--only portions of the total bound entanglement could be  displayed, it would seem.

 Along somewhat similar lines, Gabuldin and Mandilara concluded that the particular bound-entangled states they found in certain analyses of theirs had ``negligible volume and that these form tiny `islands' sporadically distributed over the surface of the polytope of separable states" \cite{gabdulin2019investigating}. In a continuous variable study \cite{diguglielmo2011experimental}, ``the tiny regions in parameter space where bound entanglement does exist'' were noted.
 
 Numerous examples of classes of bound-entangled states have appeared in the copious, multifaceted literature on the subject. It would be of interest to investigate whether or not the archipelago phenomena reported here and in \cite{slater2019bound} occur in those settings, as well.
 
 We hope to extend our studies to further models, by analyzing their entanglement properties in the interesting framework--involving the solution of the multiplicative Horn problem \cite{bercovici2015characterization}--advanced by Li and Qiao (cf. \cite{li2018necessary,gamel2016entangled}). The appropriate entanglement constraints would need to be constructed (cf. \cite{LiQiaoQuestion}). 
 
 It, then, further occurred to us that the qubit-ququart model of Li and Qiao (eq. (\ref{rhoAB})) could be directly modified, within their framework, to a $4 \times 4$ two-ququart  one of the form,
 \begin{equation} \label{2ququarts}
\rho_{AB}^{(2)}=\frac{1}{2 \cdot 8} \textbf{1} \otimes \textbf{1} +\frac{1}{4} (t_1 \lambda_1 \otimes \lambda_1+t_2 \lambda_{13} \otimes \lambda_{13}+t_3 \lambda_3 \otimes \lambda_3),
\end{equation}
where as before the $\lambda$'s are $SU(4)$ generators.
 Then, we have the corresponding 
 (independent) entanglement constraints (cf. eq. (\ref{eq3})),
 \begin{equation} \label{eq4}
  (|t_1|+|t_2|+|t_3|)^2>1 \hspace{.1in} \mbox{or}   \hspace{.15in} (t_1 t_2 t_3)^2 >  \Big(\frac{2}{27}\Big)^4,
\end{equation}
with the set of all two-ququart states being defined by the constraint
\begin{equation}
 -\frac{1}{4}<t_2<\frac{1}{4}\land -\frac{1}{4}<t_1<\frac{1}{4}\land
   -\frac{1}{4}<t_3<\frac{1}{4}.  
\end{equation}
That is, the set of possible $\{t_1,t_2,t_3\}$ comprises the cube $[-\frac{1}{4},\frac{1}{4}]^3$.
All these states have positive partial transposes.

\begin{figure}
    \centering
    \includegraphics{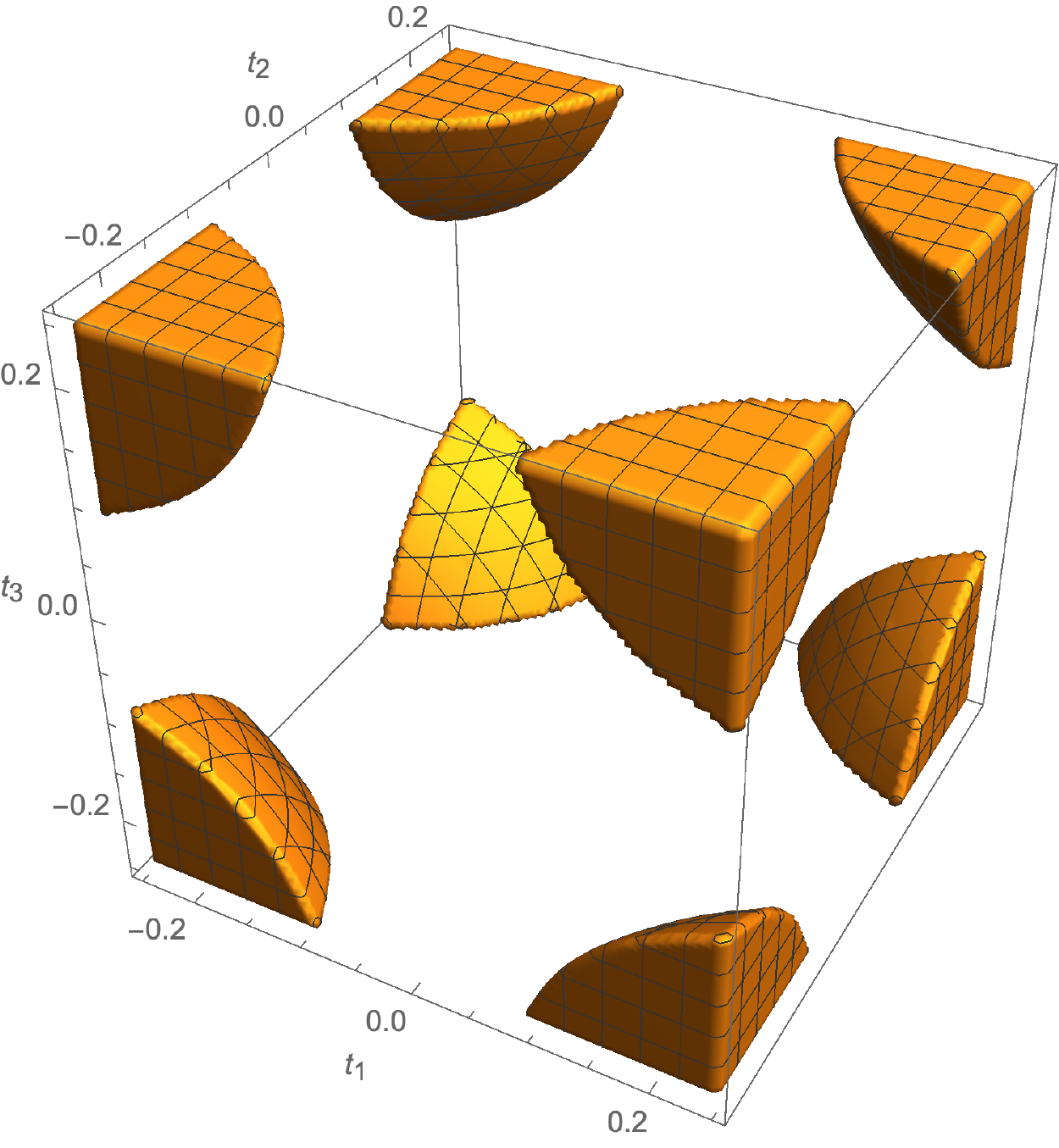}
    \caption{Bound-entangled archipelago for 
    $4 \times 4$ dimensional  two-ququart state  (\ref{2ququarts}). The total probability ($\approx 0.0890496$) captured by the islands is given by eq. (\ref{eq5}).}
    \label{fig:TwoQuQuarts}
\end{figure}

The first of the two constraints in (\ref{eq4}) again proves unenforceable--of no utility in determining the presence of any entanglement.
However, with the use of the second (multiplicative) constraint, we arrive at a bound entanglement probability simply equal to
\begin{equation} \label{eq5}
 \frac{1}{729} \left(473-512 \log \left(\frac{27}{16}\right) \left(1+\log
   \left(\frac{27}{16}\right)\right)\right) \approx 0.0890496   
\end{equation}
(all the integers being powers of 2 or 3, except that $473 =11 \cdot 43$).
Again, we find an archipelago composed of eight islands (Fig.~\ref{fig:TwoQuQuarts}), the total probabilities of which sum to this elegant result.

Further, let us now  ``downgrade'' the Li-Qiao qubit-ququart model to  simply a two-qubit one,
\begin{equation} \label{rhoABtwoqubit}
\rho_{AB}^{(3)}=\frac{1}{2 \cdot 2} \textbf{1} \otimes \textbf{1} +\frac{1}{4} (t_1 \sigma_1 \otimes \sigma_1+t_2 \sigma_2 \otimes \sigma_{13}+t_3 \sigma_3 \otimes \sigma_3),
\end{equation}
while employing the entanglement constraints,
\begin{equation} \label{eq8}
  (|t_1|+|t_2|+|t_3|)^2>1 \hspace{.1in}  \mbox{or} \hspace{.1in} (t_1 t_2 t_3)^2 > \Big(\frac{1}{27}\Big)^2.
\end{equation}
Then, we obtain a number of interesting results. Firstly, now only one-half of the physically possible states have positive partial transposes.

\begin{figure}
    \centering
    \includegraphics{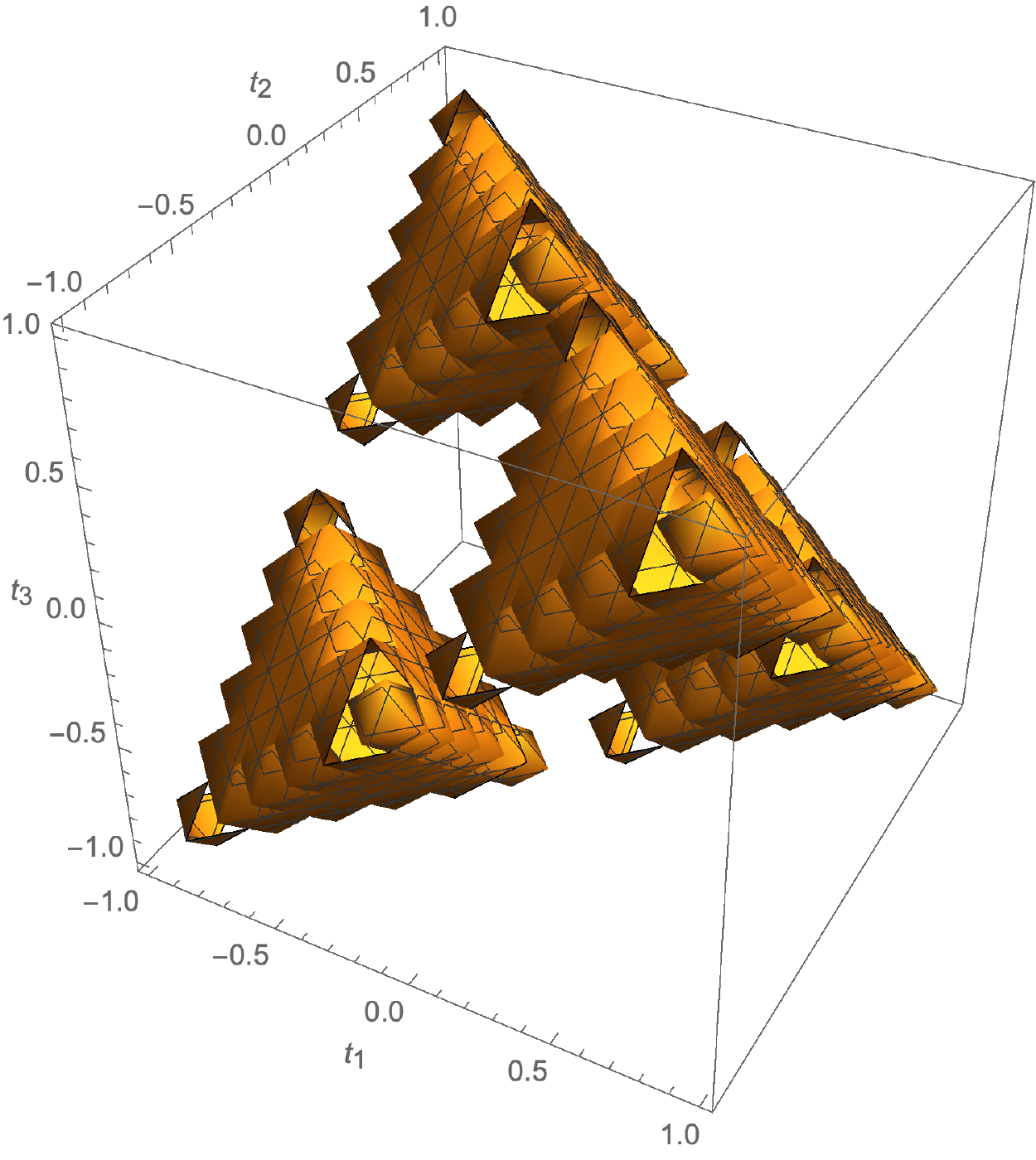}
    \caption{Archipelago of (non-bound/free) entangled two-qubit states for the set of states given by (\ref{rhoABtwoqubit}). The total probability is $\frac{1}{2}$.}
    \label{fig:TwoQubitsLiQiao}
\end{figure}

\begin{figure}
    \centering
    \includegraphics{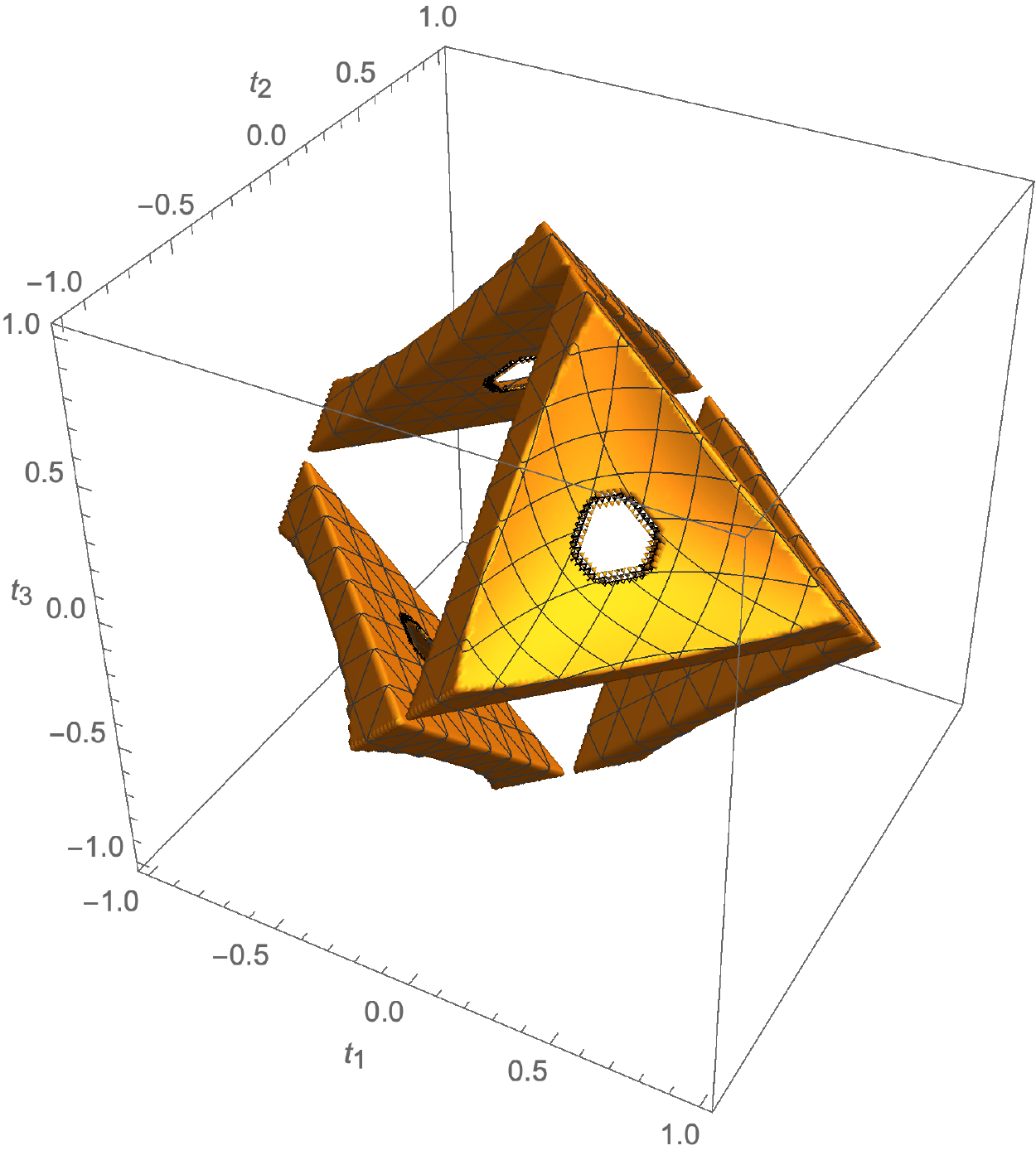}
    \caption{Those two-qubit states which satisfy the $(|t_1|+|t_2|+|t_3|)^2>1 $ entanglement  constraint, but not the $(t_1 t_2 t_3)^2 > \Big(\frac{1}{27}\Big)^2$ one. The associated probability is $\frac{1}{2}-0.3911856 =0.108814$.}
    \label{fig:PartFree}
\end{figure}

Also, imposition of the single (additive) constraint  $(|t_1|+|t_2|+|t_3|)^2>1 $ reveals that the other (non-PPT) half 
of the states are all entangled, as expected. On the other hand, enforcement of the single (multiplicative) constraint reveals that  only 0.3911855600402 of these non-PPT states are entangled. The entangled states again forms an archipelago (Fig.~\ref{fig:TwoQubitsLiQiao}), also apparently ``jagged'' in nature, but now clearly not of a bound-entangled nature (given the two-qubit context).
Those two-qubit states which satisfy the $(|t_1|+|t_2|+|t_3|)^2>1 $ entanglement  constraint, but not the $(t_1 t_2 t_3)^2 > \Big(\frac{1}{27}\Big)^2$, one are displayed in Fig.~\ref{fig:PartFree}. The associated probability is $\frac{1}{2}-0.3911856 =0.108814$.

Continuing with our analyses, we have been able to determine that the appropriate (multiplicative) entanglement constraint to employ for the first member, 
\begin{equation} \label{firstmember}
\rho_1= \frac{1}{9} \textbf{1} \otimes \textbf{1} + \frac{1}{4}(t_1\lambda_{1}\otimes \lambda_{1} +t_2 \lambda_{2}\otimes \lambda_{2}+ t_3\lambda_{3}\otimes \lambda_{3} ) \end{equation}
of the pair of
two-qutrit (octahedral and tetrahedral) models of Li and Qiao \cite[sec.2.3.2]{li2018separable} is
\begin{equation}
( t_1 t_2 t_3 )^2> \frac{2^{12}}{3^{18}}=\frac{4096}{387420489},  
\end{equation}
and for the second member, 
\begin{equation} \label{secondmember}
\rho_2= \frac{1}{9} \textbf{1} \otimes \textbf{1} + \frac{1}{4}(t_1\lambda_{1}\otimes \lambda_{1} + t_2\lambda_{2}\otimes \lambda_{4}+ t_3\lambda_{3}\otimes \lambda_{6})     
\end{equation}
of the pair,
\begin{equation}
 (t_1 t_2 t_3 )^2> \frac{2^{12}}{3^{15} } =\frac{4096}{14348907}.
\end{equation}
(We achieved these results by maximizing the product $t_1 t_2 t_3$,
subject to the conditions that the parameterized target density matrix and its separable components  not lose their positive definiteness properties.)

For the first two-qutrit model (\ref{firstmember}), we remarkably found the exact same entanglement behavior/probabilities ($\frac{1}{2}$ and 0.3911855600402 and Fig.~\ref{fig:PartFree}) as we did in  the two-qubit analyses.
Also, we did  not find that the second two-qutrit model (\ref{secondmember}) evinced any entanglement at all--in accordance with the explicit assertion of Li and Qiao that the state ``is separable for all values of $t_i$,\ldots"

Following and building upon the work of Li and Qiao, all the analyses 
reported above have involved the {\it three} parameters $t_1, t_2, t_3$, thus, lending results to immediate visualization. In higher-dimensional studies, one would have to resort to cross-sectional examinations, such as Figs. 22 and 23 in \cite{slater2019bound}, based on the ({\it four} parameter) two-ququart  Hiesmayr-L{\"o}ffler ``magic simplex" model \cite{hiesmayr2014mutually}.

It now seems possible to rather readily extend the Li-Qiao framework to (higher-dimensional) bipartite systems--{\it e. g.} qutrit-ququart, qubit-ququint,\ldots other than the specific ones studied above. Of immediate interest for all such systems is the question of to what extent they have positive partial transposes. Then, issues of bound and free entanglement can be addressed.

Let us also raise the question of whether or not the Hiesmayr-L{\"o}ffler ``magic simplices"  \cite{hiesmayr2014mutually} and/or the generalized Horodecki states \cite{horodecki1998mixed} can be studied--through reparameterizations--within the Li-Qiao framework, with consequent answers as to the associated {\it total} bound entanglement probabilities.

\begin{acknowledgements}
This research was supported by the National Science Foundation under Grant No. NSF PHY-1748958.
\end{acknowledgements}

\bibliography{main}

\end{document}